\documentclass[aps,prl,twocolumn,floats,showpacs,superscriptaddress]{revtex4}
\usepackage{graphicx,epsfig}
\usepackage{times}
\usepackage{graphics,dcolumn,bm,float}
\usepackage{amssymb,amsmath,rotate,color}

\begin{document}
\unitlength 1 cm
\newcommand{\be}{\begin{equation}}
\newcommand{\ee}{\end{equation}}
\newcommand{\bearr}{\begin{eqnarray}}
\newcommand{\eearr}{\end{eqnarray}}
\newcommand{\nn}{\nonumber}
\newcommand{\vk}{\vec k}
\newcommand{\vp}{\vec p}
\newcommand{\vq}{\vec q}
\newcommand{\vkp}{\vec {k'}}
\newcommand{\vpp}{\vec {p'}}
\newcommand{\vqp}{\vec {q'}}
\newcommand{\bk}{{\bf k}}
\newcommand{\bp}{{\bf p}}
\newcommand{\bq}{{\bf q}}
\newcommand{\br}{{\bf r}}
\newcommand{\bR}{{\bf R}}
\newcommand{\up}{\uparrow}
\newcommand{\down}{\downarrow}
\newcommand{\fns}{\footnotesize}
\newcommand{\ns}{\normalsize}
\newcommand{\cdag}{c^{\dagger}}

\title{Substitutional doping of Cu in diamond: Mott physics with $p$ orbitals}
\author{H. Hassanian Arefi}
\affiliation{Department of Physics, Tarbiat Modarres University, Tehran, Iran}
\author{S. A. Jafari{\footnote {Electronic address:
akbar.jafari@gmail.com}}}
\affiliation{Department of Physics, Isfahan University of Technology, Isfahan 84156-83111, Iran}
\affiliation{School of Physics, Institute for Research in Fundamental Sciences (IPM), Tehran 19395-5531, Iran}

\author{M. R. Abolhassani}
\affiliation{Department of Physics, Tarbiat Modarres University, Tehran, Iran}
\date{\today}

\begin{abstract}
Discovery of superconductivity in the impurity band formed by heavy doping of boron into 
diamond (C:B) as well as doping of boron into silicon (Si:B) has provided a rout for the 
possibility of new families of superconducting materials. Motivated by the special role played
by copper atoms in high temperature superconducting materials where essentially Cu $d$ orbitals
are responsible for a variety of correlation induced phases, in this paper we investigate
the effect of substitutional doping of Cu into diamond. Our extensive first principle calculations  
averaged over various geometries based on density functional theory, indicates the formation 
of a mid-gap band, which mainly arises from the $t_{2g}$ and $4p$ states of Cu. 
For impurity concentrations of more than $\sim 1\%$, the effect of $2p$ bands of neighboring carbon 
atoms can be ignored. Based on our detailed analysis, we suggest a two band model for the mid-gap
states consisting of a quarter-filled hole like $t_{2g}$ band, and a half-filled band of $4p$ states.
Increasing the concentration of the Cu impurity beyond $\sim 5\%$, completely closes the spectral gap of 
the host diamond.
\end{abstract}
\maketitle

\section{Introduction}
Diamond is the best thermal conductor~\cite{Field, Davies}, and the hardest
known material with semiconducting properties. Both pristine form as well as
doped films of diamond present unique ground for high-temperature and
high-power applications in the industries~\cite{Davies,Nebel,Lawrence}.
   Recent discovery of superconductivity in heavily boron doped diamond, has revived 
the interest in impurity band formation from a fundamental point of view~\cite{Ekimov,Takano}.
Similar system of boron doped silicon was also studied~\cite{BlaseNature,Blase,BlaseAPL}.
Synthesis of boron doped diamond was first achieved by 
Ekimov and coworkers~\cite{Ekimov} and subsequently by
Takano {\em et. al.}~\cite{Takano}. They used chemical vapor deposition (CVD)
for the synthesis of B-doped diamond. At low doping concentrations of
typically $10^{17}-10^{18}$~cm$^{-3}$, boron atoms inject 
acceptor levels in the electronic energy spectrum of host diamond, hence producing a
$p$ type semiconductor~\cite{KalishBook}.  For heavier dopings
$\sim$[B]/[C]$>5000$~ppm in the gas phase, one achieves
metallic conductivity in diamond~\cite{Deneuville}. 
Eventually increasing the doping beyond $n> 10^{21}$~cm$^{-3}$, i.e. few percents, it
superconducts at low temperatures~\cite{Ekimov,Takano}. 
For higher concentrations, the dopant atoms come closer to each other
and start to overlap. Therefore the impurity levels are broadened
into authentic impurity bands~\cite{DanWu,Baskaran}, which are responsible for metallic 
and superconducting properties~\cite{Baskaran,KWLee,Pogorelov,Nakamura,Blase,Mukuda,Fukuyama}.

Various forms of defects play crucial role in the physical properties
of diamond. For example the fascinating colors of diamond as gem stone, 
is due to the $\sim $ppm concentration of vacancies~\cite{Collins}. Such
vacancies at much higher concentration of the order of few percents, can give rise 
to an impurity bands inside the diamond gap, which can provide larger number of states 
than some of the elements from the third and fifth column of 
the periodic table of elements~\cite{AlaeiJafari}. Such mid-gap bands 
of the vacant diamond can be attributed to a localized Wannier wave 
function composed from the $2p$ states of neighboring carbon atoms.
Therefore, previous studies of the possible conductive impurity bands in diamond
and/or silicon hosts, have been mainly concerned with the substitution of
a $p$-type orbital for the host $p$-states.

  In this paper, we are interested in the role of 
substitution by a $3d$ element. We choose to study the effect of copper
substitution in the diamond host. Why is Cu interesting from the impurity
band formation point of view? 
First of all, the same high-temperature, high-pressure techniques employed
for the synthesis of boron doped diamond can also be applied to attempt
a substitution of Cu atoms into the diamond host~\cite{Ghodselahi}.
From theoretical point of view, although Cu is a $3d$ transition metal, it does
not have magnetic complications, and hence one can avoid possible magnetic
ground states at low temperatures which tend to compete against superconducting correlations.
Choice of Cu also avoids the possibility of local moment formation.
On the other hand $3d^9$ "atomic" configuration of Cu is hoped to provide a hole which
plays an important role in the high temperature superconducting (HTSC) materials~\cite{Anderson}.
The nominal $3d$ hole of atomic Cu in HTSC systems is shared between neighboring oxygen $2p$ 
and Cu $3d$ states, giving rise to the so called Zhang-Rice singlet state~\cite{ZhangRice} which
carries the hole. 

   In this paper, we study the formation of impurity bands in Cu doped diamond
in detail and we find that $4p$ states of Cu along with its $t_{2g}$ orbitals
are responsible for the formation of impurity band at Cu concentrations $1-5\%$.
At lower concentrations $2p$ orbitals of neighboring carbon atoms also participate
in the formation of mid-gap metallic band.
Depending on the experimental conditions, either a combination of most probable geometries is realized, 
or by slow cooling the system will be given enough time to perform a self-averaging 
over the phase space of possible geometries.
Therefore we undertake extensive averaging over distinct geometries.
The main role of averaging is to restore the symmetries broken by the
specific realization, e.g. most probable geometry (MPG). For comparison,
we will provide MPG and geometry averaged (GA) results together.

\section{METHOD OF CALCULATION}
The calculations of electronic structure are performed with the 
plane-wave pseudopotential~\cite{Espresso} of DFT using the PBE functional~\cite{JPPerdew} 
for exchange and correlation as implemented in the QUANTUM-ESPRESSO package. 
Ultra-soft pseudopotentials are used with a $30$ Ry ($230$ Ry) 
cutoff for the expansion of the wave-functions (charge density). We have constructed 
$2\times 2\times 2$ supercell containing $64$ atoms with $1$, $2$ and $3$ Cu impurity, 
leading to doping concentrations of $1.6\%$, $3.1\%$ and $4.7\%$, respectively. The 
sampling of Brillouin zone (BZ) for structural relaxation and DOS calculations are 
performed on $2\times 2\times 2$ and $4\times 4 \times 4$ grid of $k$-points, respectively. 
We first minimized the energy 
of undoped diamond with respect to cell size to calculate the lattice constant. 
The obtained value was very close to the experimental result. All structures were 
relaxed by BFGS algorithm~\cite{Fletcher}.
 
 Two types of doping can be considered: substitutional and interstitial. 
In the case of B-doped diamond, interstitial doping is unstable~\cite{Oguchi}. 
Since our study is motivated by B-doped diamond,  we only consider substitutional doping 
of copper. We randomly substituted Cu impurity into the diamond structure. Since there
are many different ways of introducing impurity into the host carbon lattice, one needs 
to average over sufficient number of realizations of randomness. With $m$ impurities 
in a host of $n$ lattice points, there 
are $C(n,m)=\frac{n!}{(n-m)!m!}$ different ways of substitution. 
With $n=64$ carbon atoms considered here, the most trivial case
corresponds to $m=1$ impurity. By translational symmetry (periodic boundary conditions), 
it does not matter where we place the 
single impurity. Therefore the average DOS corresponding to $1$ impurity, is identical 
to any of the random realizations.
Next in the case of $m=2$ impurities, there are in principle $C(64,2)=2016$ ways of 
placing the Cu impurities. However, since the 
{\em relative distance}  between the impurities is the physically discriminating 
factor, many of the above configurations will be essentially identical by translation
and rotation symmetries. Therefore the electronic structure calculations must be run for 
geometrically different configurations. We have used a computer program to count the number of 
geometrically independent configurations which in the case of $m=2$ impurities
gives $23$. Therefore one has to perform the electronic structure for all $23$ 
distinct geometries, taking into account the multiplicity $w_g$ of each geometry $g$.
In the case of $m=3$ impurities, we have $C(64,3)=41664$ nominally different configurations. 
Again our counting algorithm gives
$430$ distinct geometries, where each geometry is meant to 
represents a distinct set of $3$ mutual distances between the impurity atoms.
We used $60$ most probable geometries to calculate the averages. 
Note that the above $60$ geometries cover more than half of the total 
$41664$ choices. We have checked that, the average DOS obtained for $60$ most probable 
geometries can be more or less achieved even with $10$ most likely geometries.
The geometry averaged values of various quantities $X$ such as $\rho(\omega)$, $E_F$ 
can be calculated as follows:
\be
    X_{\rm GA}=\sum_{g} w_g X_g,
\ee
where $g$ runs over geometries, and $w_g$ is its relative weight.

\begin{figure}[t]
  \begin{center}
    \includegraphics[scale=0.5,angle=-90]{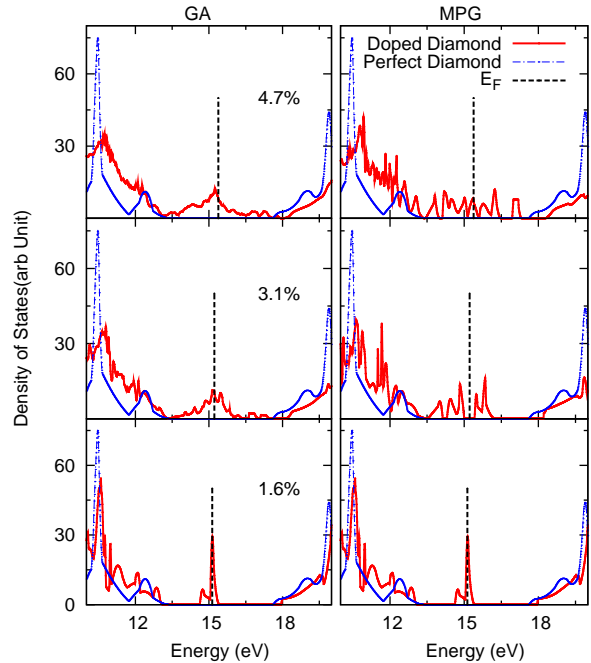}
    \caption{(Color online) Total density of states for three considered concentrations contrasted against 
    pure diamond in background. Left and right panels correspond to geometry averaged and most probable
    geometries, respectively.}
    \label{DOS.fig}
  \end{center}
\end{figure}

\section{Metalization of diamond}
In Fig.~\ref{DOS.fig} we have plotted the DOS of doped diamond lattice with (left) and 
without (right) averaging over geometries.
The doping rates  in the figure correspond to $1$, $2$, $3$ Cu impurities 
in a host of $64$ carbon atoms, from bottom to top, respectively. 
The presence of Cu impurities, affects the
entire spectrum (not shown here). In this figure we have focused on the region of 
spectrum corresponding to gap in the pristine diamond. 
By increasing the doping rate from $1.6$ to $4.7$ percent, 
the deviation in the DOS of Cu-substituted diamond from 
the DOS of perfect diamond becomes more and more significant. 

The most spectacular effect of substituting Cu in the insulating host of
diamond, is the formation of mid-gap impurity band. In the right panel of Fig.~\ref{DOS.fig},
DOS corresponding to MPG is shown. By averaging over 
different geometries, the corresponding geometry averaged DOS in the left panel will be
obtained. As can be seen in the figure, for both GA and MPG, 
the width of the impurity band increases with increasing the Cu concentration.
At $4.7\%$ Cu substitution, the gap completely disappears. The averaged Fermi
energy, $E_F$ is indicated with vertical dashed line. As it is evident, 
although the most probable geometry in the case of $2/64$ impurity concentration
gives a little gap in the spectrum, other geometries with high probability
will fill in that small gap and the resulting GA density of states will 
be a metallic band. This indicates that the best 
known insulator, namely diamond, can be metalized by few percent Cu substitution.
The question we would like to investigate in detail in this paper is,
what is the origin of the metallic band?

\begin{figure}[t]
  \begin{center}
    \includegraphics[scale=0.53,angle=-90]{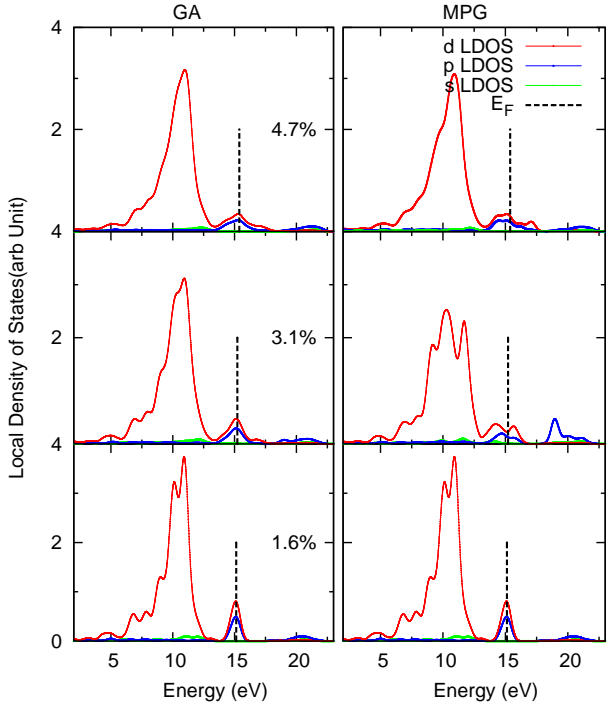}
    \caption{(Color online) LDOS for $s$, $p$ and $d$ orbitals of Cu corresponding
    to three concentrations considered. Fermi level $E_F$ is indicated by vertical
    dashed line. MPG values on the right panel are given for comparison.
    As can be seen the contribution of $4s$ orbitals
    to the impurity band is negligible.}
    \label{Cu-LDOS.fig}
  \end{center}
\end{figure}
 
\subsection{Role of Cu atoms}
   To consider this question, we first examine the occupation of various orbitals. 
We find that unlike the atomic configuration $4s^2 3d^9$ of Cu, in the presence of 
carbon neighbors, the occupancy of $4s$ orbitals is changed as $4s^2\to4s^{0.5}$. 
The missing $1.5$ electron changes the $3d^9\to3d^{9.5}$, $4p^0\to 4p^1$. 
Hence $4p$ orbitals are pushed down to acquire occupancy of $\sim 1$. We find that
the above Lowdin charges are more or less {\em independent of the geometry,
as well as concentration of Cu impurities}. This can be qualitatively seen in 
Fig.~\ref{Cu-LDOS.fig}, where local density of states (LDOS) for $s$, $p$ and $d$ 
orbitals of Cu are plotted.  Therefore the relevant orbitals of Cu are both $4p$, 
as well as a subspace of $3d$ states. 

\begin{figure}[t]
  \begin{center}
    \includegraphics[scale=0.5,angle=-90]{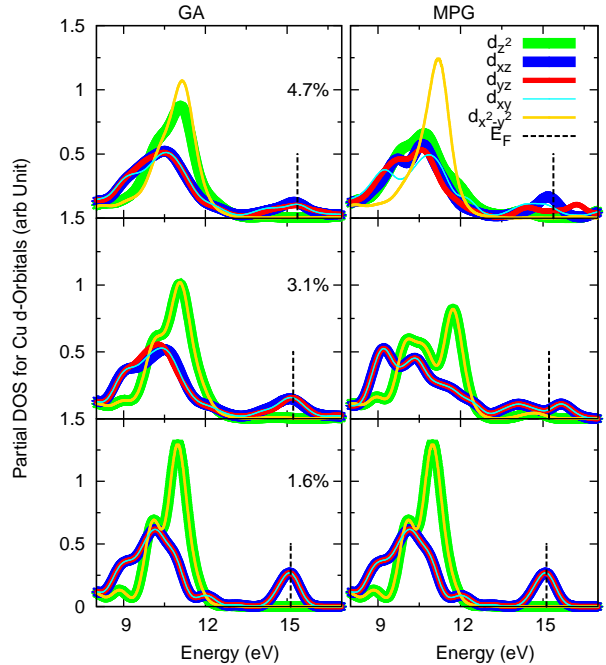}
    \caption{(Color online) Copper $d$-orbital splitting for three considered concentrations 
    in diamond crystalline field. The major contribution to the impurity band comes
    from $t_{2g}$ subspace of $d$-orbitals. }
    \label{d-splitting.fig}
  \end{center}
\end{figure}

To further identify which subspace of Cu $d-$orbitals contribute to impurity band,
in Fig.~\ref{d-splitting.fig} we resolve the contribution of $t_{2g}$ and $e_g$
sub-bands. In the case of one Cu atom, the impurity atom feels a perfect
diamond environment; thereby giving three-fold degenerate $t_{2g}$ band and
two-fold $e_g$ bands. Note that $e_g$ bands are located at lower energies
with respect to $t_{2g}$ bands. By increasing the number of Cu atoms,
the above crystalline degeneracies start to be lifted, as the crystalline environment
at higher Cu concentrations will deviate from perfect diamond symmetry. 
At the same time the average distance between Cu atoms decreases. 
Especially in the case of
three Cu atoms, diamond crystalline environment is totally lost in the DOS of MPG.
Even after averaging over geometries, substantial degeneracy lifting in each 
$t_{2g}$ and $e_g$ sub-bands can be observed. The important result
from Fig.~\ref{d-splitting.fig} is that in all three considered concentrations, 
and for both MPG and GA situations, only $t_{2g}$ sub-band of the $d$-orbitals
contribute to the impurity band, while the $e_g$ sub-band will give energy levels
inside the valence band. 

\subsection{Role of neighboring carbon atoms}
\begin{figure}[t]
  \begin{center}
    \includegraphics[scale=0.5,angle=-90]{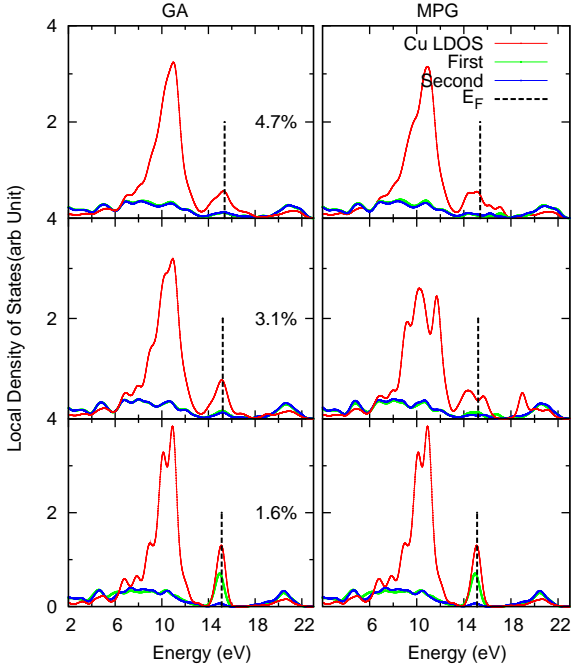}
    \caption{(Color online) Spatially resolved DOS at the Cu site. Right panel 
    corresponds to MPG, and the left panel includes averaging over geometries.
    At high concentrations the dominant contribution comes from Cu atom itself,
    while at lower concentration nearest neighbor carbon atoms have substantial
    contribution to the impurity band. 
    }
    \label{neighbors.fig}
  \end{center}
\end{figure}

  Now let us concentrate on the effect of neighboring carbon atoms on 
the formation of the impurity band. In Fig.~\ref{neighbors.fig} we plot
DOS corresponding to three concentrations for MPG and GA. 
In terms of the relative weights of neighbors, both MPG and GA give 
similar results. Note that for one Cu impurity, MPG and GA quantities must 
always be the same. By moving from bottom to top (increasing the concentration)
in Fig.~\ref{neighbors.fig} one notices that only at low concentration 
nearest neighbor (NN) carbon atoms contribute significantly.
By increasing the concentration, contribution of NN carbon atoms compared 
to the Cu atoms is reduced and becomes as low as the next nearest neighbor (NNN) 
contributions. This can be attributed to the fact that at higher Cu concentration,
average Cu-Cu distance is reduced and the Cu-Cu hopping matrix elements become
dominant. At lower concentrations in addition to Cu-Cu hoppings, there is also
substantial hopping amplitude between the dangling  orbitals of neighboring carbon 
atoms. This can be thought of hopping between appropriate Wannier orbitals localized 
around Cu atoms which is composed of the dangling $2p$ orbitals of neighboring carbon
atoms. 

\begin{figure}[t]
  \begin{center}
    \includegraphics[scale=0.5,angle=-90]{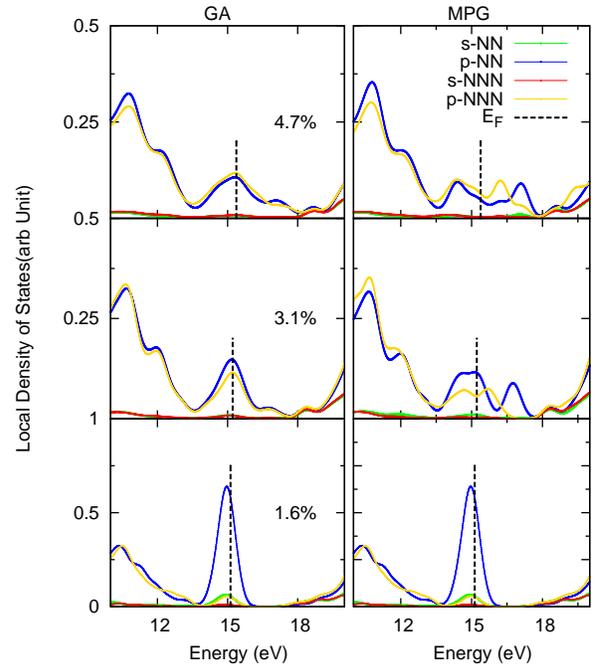}
    \caption{(Color online) Contribution of first and second neighbors (NN and NNN)
    to the DOS of the impurity band. Right panel correspond to MPG and the left panel
    includes averaging over geometries. As can be seen $s$ orbitals of neighboring carbon
    atoms do not contribute to the impurity band. The relative contribution of neighboring 
    $p$ orbitals to the Cu atomic orbitals become significant in low concentration limit.
    }
    \label{neighbors-s-p.fig}
  \end{center}
\end{figure}

Based on a naive $sp^3$ picture of bonding in perfect diamond, one expects $2s$
and $2p$ orbitals of neighboring carbon atoms to have comparable contribution
to the formation of impurity bands. However, Cu substitution drastically 
changes this picture. As can be seen in Fig.~\ref{neighbors-s-p.fig},
in both MPG and GA situations $s$ orbitals are irrelevant and the 
contribution of NN and NNN carbon atoms comes from their $2p$ states.
These contributions become comparable to local DOS arising from Cu atoms only
in low Cu concentration limit. When comparing 
Figs.~\ref{neighbors.fig},~\ref{neighbors-s-p.fig} one has to note the scales.
Note that 1/64 concentration can also be reproduced in the simulation with two 
Cu atoms in host supercell of $128$ carbon sites. Therefore the growth in the relative 
weight of neighboring $p$ orbitals can be partly due to finite size artifacts of
the simulation, which of course is expected to be minimal with the periodic boundary
conditions employed in the calculations. Therefore we expect the neighboring atoms to have 
significant $2p$ orbital contributions for concentrations well below $\sim 1\%$. 
At higher concentrations, the contribution of neighboring atoms 
becomes less and less significant. A comparison of MPG and GA density of states in 
Fig.~\ref{neighbors-s-p.fig} indicates that the role of averaging
is to slightly increase the contribution of $p$-orbitals which is
essentially a statistical effect. 

\section{Discussion and conclusion}
Examining the bond lengths show that  Cu-C bond is elongated by approximately 
$0.2$ relative to C-C bonds in undoped structure which has been almost compensated 
by a similar decrease in the surrounding C-C bond lengths. 
C-C bonds far away from Cu undergo negligible changes of the order
$0.01\AA$. Therefore in terms of structural changes, Cu substitution will give rise to
local effects. In terms of electronic properties, in the regime of $\sim 5\%$ Cu substitution,
the most significant process behind the formation of impurity band, 
is the hopping between Cu atoms themselves.
In this regime a minimal two-band effective model can describe the electronic states near
the Fermi energy:
\bearr
   H &=&t_{dd}\sum_{\langle i,j\rangle \sigma}
   \left(d^\dagger_{i\sigma}d_{j\sigma}+\mbox{h.c.}\right)
   -t_{pp}\sum_{\langle i,j\rangle \sigma}
   \left(p^\dagger_{i\sigma}p_{j\sigma}+\mbox{h.c.}\right)\nn\\
   &&+t_{pd}\sum_{\langle i,j\rangle \sigma}
   \left(d^\dagger_{i\sigma}p_{j\sigma}+\mbox{h.c.}\right),
   \label{mymodel.eqn}
\eearr
where the hopping matrix elements $t_{dd},t_{pp}>0$ are between the 
$t_{2g}$ {\em hole} bands and $4p$ {\em electron} bands of the adjacent Cu 
atoms residing at random sites $i,j$. If the inter-band matrix element $t_{pd}$
is negligible, one would have a quarter filled band of $3d$ holes, along with 
a half-filled $4p$ band constituting the impurity band. In this limit 
for large enough Coulomb correlations among the Wannier orbitals derived from 
parent $4p$ states, the electron-like band undergoes a Mott transition, while
the 1/4-filled hole bands derived from $t_{2g}$ states can be subject to 
various forms of density waves~\cite{FukuyamaRev}. Theoretically speaking, in the opposite limit 
when  $t_{pd}$ is large, the original half-filled electron band and the quarter 
filled hole band, get mixed giving rise to an effective quarter filled band
whose electrons have a mixed character,
\be
    c^\dagger_{i\sigma}=\frac{d^\dagger_{i\sigma}+p^\dagger_{i\sigma}}{\sqrt 2}.
\ee
This can be thought of as analogue of Zhang-Rice singlet state~\cite{ZhangRice}.
Although this limit seems to be  hard to realize in the experiment, but from
theoretical point of view it may prove useful in studying  transition from
two-band situation to effective one-band model.
In the intermediate regime where $|t_{pd}|\sim t_{pp},t_{dd}$, copper doped
diamond will offer an opportunity to study both theoretically and experimentally
the competition between the instabilities of a half-filled electronic band~\cite{Fujimori} with that 
of a quarter-filled hole band~\cite{FukuyamaRev}. Our model Hamiltonian~(\ref{mymodel.eqn}) defines the
competition between instabilities of two bands in 3D. The randomness in 
the matrix elements can not be avoided, as there is no control over the location 
of Cu impurities. Robustness of Lowdin charges in various geometries imposed 
the restriction $t_{dd},t_{pp}>0$ for studying Anderson localization effects in 
such 3D bands. The fact that in $t_{pd}\ll t_{pp},t_{dd}$ limit electronic bands
of $4p$ orbitals are half-filled, suggests the possibility of Mott-Anderson physics
in $4p$ orbitals. 

To summarize, from both experimental and theoretical point of view, impurity
band formation in Cu/C system provides interesting ground to study the 
orbital selective phenomena and novel forms of Anderson localization properties 
in a combination of electron and hole like bands with different fillings in 3D.

\section{Acknowledgement}
We thank Dr. M. Alaei for fruitful discussions. 
H.H.F. would likes to thank R. Sarparast, K. Haghighi and M. Hafez for 
assistance in computing.  S.A.J. was supported by the Vice Chancellor for
Research Affairs of the Isfahan University of Technology (IUT)
and the National Elite Foundation (NEF) of Iran.

\end{document}